\documentclass{ws-rv9x6}
\usepackage{ws-rv-van}             

\usepackage{bm}
\usepackage{ws-index}             
\makeindex
\newindex{aindx}{adx}{and}{Author Index}       
\renewindex{default}{idx}{ind}{Subject Index}  

\begin{document}

\chapter{Single-Electron Tunneling and the Fluctuation Theorem}

\author{ Y. Utsumi$^1$, D. S. Golubev$^2$, M. Marthaler$^{3}$, 
T. Fujisawa$^{4,5}$, 
and Gerd~Sch\"on$^{2,3}$ }

\address{
$^1$Institute for Solid State Physics, University of Tokyo, Kashiwa, Chiba 277-8581, Japan \\
$^2$ Institut f\"ur Nanotechnologie, Karlsruhe Institute of Technology, 76021 Karlsruhe, Germany \\
$^3$Institut f\"{u}r Theoretische Festk\"{o}rperphysik and DFG Center for Functional Nanostructures (CFN), Karlsruhe Institute of Technology, 76128 Karlsruhe, Germany \\
$^4$NTT Basic Research Laboratories, NTT Corporation, Morinosato-Wakamiya, Atsugi 243-0198, Japan \\
$^5$Research Center for Low Temperature Physics, Tokyo Institute of Technology, Ookayama, Meguro, Tokyo 152-8551, Japan 
}

\begin{abstract}
Experiments on the direction-resolved full-counting statistics of single-electron tunneling allow testing
the fundamentally important Fluctuation Theorem (FT). At the same time, the FT provides a frame
for analyzing such data. Here we consider tunneling through a double quantum dot system which is 
coupled capacitively to a quantum point contact (QPC) detector. 
Fluctuations of the environment, including the shot noise of the QPC, lead to 
an enhancement of the effective temperature in the FT. We provide a quantitative explanation of this effect; in addition we discuss the influence of the finite detector bandwidth on the measurements.
\end{abstract}

\body

\section{Introduction}
The second law of thermodynamics states that the entropy
of a macroscopic system driven out of equilibrium
grows with time, and the dynamics of such a system is
irreversible. 
The entropy of a mesoscopic system also grows in the long-time limit, but it
may decrease over sufficiently short periods of time. 
Hence, the entropy production $\Delta S$ during  a time interval $\tau$ is a random variable, characterized by a distribution $P_\tau(\Delta S)$. 
The `Fluctuation Theorem' (FT)~\cite{Evans} states that the probabilities
of positive and negative entropy changes at sufficiently 
long $\tau$ are related   by
\begin{equation}
\frac{P_\tau(\Delta S)}{P_\tau(-\Delta S)}=\exp(\Delta S).
\label{DS}
\end{equation}
Remarkably, this simple and universal relation remains valid even far from equilibrium. 
It has been proven for thermostated Hamiltonian systems~\cite{Evans}, 
Markovian stochastic processes~\cite{Lebowitz,Andrieux,Andrieux2}, 
quantum systems~\cite{Kurchan}, 
and 
 mesoscopic conductors~\cite{Tobiska,FB,SU,US,Esposito,Andrieux1}. 
The FT is fundamentally important for transport theory.
One of its consequences is the Jarzynski equality~\cite{Jarzynski,Campisi},
which in turn leads to the 2nd law of thermodynamics. 
It also leads to the fluctuation-dissipation theorem and Onsager symmetry relations \cite{Gallavotti}, 
as well as to their extensions to nonlinear transport~\cite{Tobiska,FB,SU,US,Esposito,Andrieux1}. 

The FT was first verified  in an 
experiment measuring the distribution of the work done on a colloidal particle
 placed in a water flow and  trapped by an optical tweezer~\cite{Wang}. 
By monitoring the position fluctuations it is possible to estimate 
the work done on the particle. For this classical experiment, as well as for other related ones
performed at room temperature~\cite{Liphardt}, the thermal fluctuations are quite large and   
the FT has been confirmed. 
In contrast, experiments for mesoscopic quantum systems~\cite{Imry} were lacking until very recently~\cite{our_paper,Nakamura}. 

Let us now discuss the implications of FT for mesoscopic systems. 
The first experimental test of the FT applied to single-electron transport has been performed 
recently in Ref.~\cite{Fujisawa,our_paper}. 
A system of two coupled quantum dots in a 2DEG at the GaAs/AlGaAs interface
was operated in the Coulomb blockade regime. 
(The single-electron charging energy of a single dot was of the order of 100 $\mu$eV,
and the sample was cooled to 100 milli-Kelvin.)  The single-electron tunneling 
through the double dot system was detected via the current through a nearby quantum point contact (QPC). 
The time resolution of the readout was better than 0.1 ms. 
By using an asymmetric setup the direction of tunneling could be resolved, 
and the probability distribution of forward and backward tunneling processes could be determined.

The entropy production in this experiment
is related to Joule heating and reads $\Delta S= qeV_S/ k_{\rm B} T$, 
where $q$ is the number of electrons (with charge  $e$) transfered through the conductor during time $\tau$
and $V_S$ is the bias voltage.
Hence the FT can be formulated in terms of the distribution of transfered charge  
$P_\tau(q)$ at sufficiently long times, $\tau\gtrsim e/I$, where $I$ is the current, as follows
\begin{equation}
\frac{P_\tau(q)}{P_\tau(-q)}=\exp\left(\frac{qeV_{\rm S}}{k_{\rm B} T}\right). 
\label{pft}
\end{equation} 

Unlike in classical systems, the fluctuations of the charge transfered through the quantum
dots are strongly affected by the environment (phonons and electromagnetic environment) 
and by the measurement backaction. 
In what follows we discuss various environmental effects and demonstrate their importance
for the interpretation of the experiment. We show that the non-equilibrium 
electromagnetic fluctuations caused by the non-equilibrium shot noise of the QPC detector lead to an apparent violation of the FT. 
However, we find the FT to be satisfied if we replace the temperature $T$ in Eq. (\ref{pft}) 
by an enhanced effective temperature $T^*$, which we relate to the tunneling rates for the various relevant processes. 
We also study the effect of finite bandwidth of the detector~\cite{naaman,Gustavsson2,flindt}
and show that in the parameter regime of our experiment,
this effect may also be accounted for by an effective temperature. 

The paper is organized as follows.
In Sec.~\ref{sec:et}, we briefly discuss the experimental results. 
In Sec.~\ref{sec:fts}, we discuss the theory relevant for our experiment assuming a perfect detector. 
In Sec.~\ref{sec:be}, we discuss the effect of the environment. 
The realistic case, where the time resolution of the detector
is limited by its bandwidth, will be discussed in  Sec.~\ref{sec:finite}.
There we will show that it is still possible to recover the 
FT with properly corrected tunneling rates. 
In Sec.~\ref{sec:wtd}, we show how the tunneling rates are determined experimentally and how the finite bandwidth of the detector affects the experimentally obtained value. 
Section~\ref{sec:sum} summarizes our discussion.

\section{Experimental test of the FT in single-electron counting}
\label{sec:et}

\begin{figure}[hb]
\begin{center}
\includegraphics[width=0.9 \columnwidth]{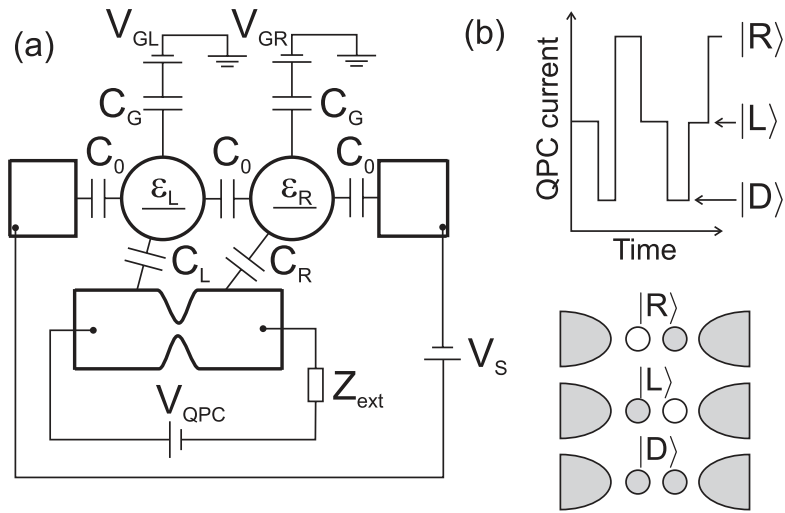} 
\caption{
(a) Setup of the system with two quantum dots (DQD) with single-level energies $\varepsilon_L$ and $\varepsilon_R$ coupled to a quantum point contact (QPC). 
(b) The QPC current switches between three values corresponding to the three charge states of the DQD. 
} 
\label{Figsetups}
\end{center}
\end{figure}

 Most of the experiments on single-electron counting are performed with a single quantum dot capacitively coupled to a QPC detector \cite{ensslin_review}. In this case the QPC current switches between the two values corresponding to an occupied and an empty quantum dot. 
Such a detector cannot resolve the direction of the electron tunneling and, therefore, is not suitable for testing the FT. 
The simplest system which does resolve the direction of the tunneling consists of two serially coupled quantum dots which are asymmetrically coupled to a QPC detector~\cite{Fujisawa} (Fig. \ref{Figsetups} a). 
The left and right gate voltages, $V_{\rm GL}$ and $V_{\rm GR}$, applied 
to the quantum dots are tuned in such a way that only three charge states of the DQD need to be considered.   
In the experiment \cite{Fujisawa} those states are $| D \rangle$ (both dots are occupied), $|L\rangle$ 
(left dot is occupied by one electron) and $|R\rangle$ (right dot is occupied). 
Accordingly, the current through the QPC, which is coupled asymmetrically to the DQD, switches between three different values (Fig.~\ref{Figsetups} b). 
This setup allows distinguishing electron tunneling in different directions and between the dots and leads.

\begin{figure}
\begin{center}
\includegraphics[width=0.7 \columnwidth]{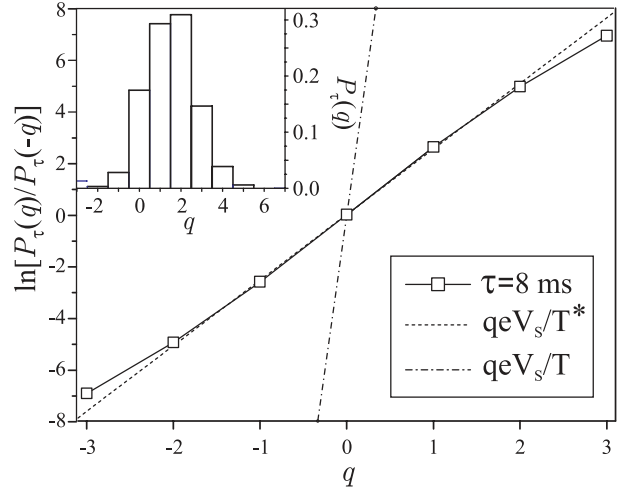} 
\caption{
Test of FT (\ref{pft}) with the measurement time $\tau=8$ ms. 
Lines with squires: logarithm of lhs of Eq. (\ref{pft});
dashed line: 
$q\, eV_S/ k_{\rm B}T^*$  with $T^*=1.37$ K; 
dot-dashed line: $q\, eV_S/ k_{\rm B}T$. 
Inset: the distribution $P_\tau(q)$ at $\tau=4$ ms. 
} 
\label{Figexptest}
\end{center}
\end{figure}

From the time trace of the current taken during time $\tau$ one obtains 
the distribution of transfered charges between the two dots, $P_\tau(q)$, an example of which is shown in the inset of Fig.~\ref{Figexptest}. 
In Fig.~\ref{Figexptest} we perform a test of the FT (\ref{pft}).
The combination $\ln[P_\tau(q)/P_\tau(-q)]$ depends indeed linearly on
the transfered charge $q$ with a slope $eV_S/ k_{\rm B} T^*$. Here $V_{\rm S} = 300$ $\mu$V is the applied DQD bias voltage, but the effective temperature $T^*=1.37$ K fitting the data  (dashed line) strongly exceeds the bath temperature of the leads of $T=130$ mK (dot-dashed line).

\begin{figure}
\begin{center}
\includegraphics[width=0.5 \columnwidth]{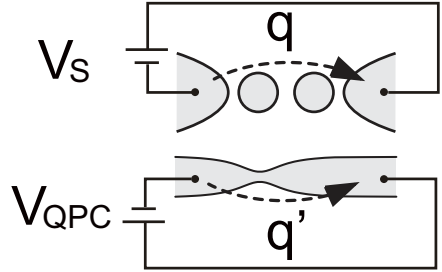} 
\caption{
Schematic picture of the DQD capacitively coupled to the QPC illustrating
the apparent violation of the FT. 
In addition to the source-drain bias $V_S$ applied to the QDQ  
the voltage $V_{\rm QPC}$ is applied to the QPC. 
Thus, the total system is a 4-terminal setup,
and the FT should be formulated for the joint
probability distribution of transmitted charges through the QPC and DQD, $P(q,q')$. 
The FT for $P(q)$ is valid only for $V_{\rm QPC}=0$. 
} 
\label{Figdqdqpc}
\end{center}
\end{figure}

In order to understand this apparent violation, we have to consider the total system. 
The FT for the system composed of the DQD and QPC should be formulated in terms of the joint probability distribution $P_\tau(q,q')$, where $q$ and $q'$  charges are transmitted through the DQD and the QPC, respectively (Fig.~\ref{Figdqdqpc})~\cite{SU}.
It satisfies 
\begin{eqnarray}
P_\tau(q,q') = 
\exp \left(
\frac{
q e V_{\rm S}+q' e V_{\rm QPC}
}{k_{\rm B} T}
\right) P_\tau(-q,-q'). 
\label{P2}
\end{eqnarray}
Since only the number of charges $q$ is measured, Eq.~(\ref{P2}) should be summed over $q'$. 
As a result, the right hand side deviates from $\exp[eV_{\rm S}/ k_{\rm B}T]\,
P_\tau(-q)$ (except for $V_{\rm QPC}\!=\!0$), which appears to violate the FT. 
In the following section we will provide a more detailed discussion and further the reasoning why the FT is recovered when we introduce the effective temperature. 

\section{FT in the single-electron transport}
\label{sec:fts}

\begin{figure}[th]
\begin{center}
\includegraphics[width=.3 \columnwidth]{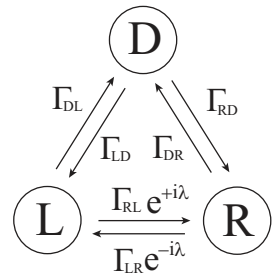}
\caption{
The relevant transition processes.  
Circles represent the double-dot states and arrows the directions of the transitions. 
The factor ${\rm e}^{\pm i \lambda}$, needed for the Full Counting Statistics, 
indicates that the electron number is `counted' at the center barrier. 
}
\label{Fig:mc1}
\end{center}
\end{figure}

 In the experiment, electrons tunnel through the dots sequentially. 
The system is then fully characterized by the vector of the occupation
probabilities of the DQD charge states, ${\bf p}^T=(p_L,p_R,p_D)$, 
which satisfies the following master equation
\begin{eqnarray}
\partial_t {\bf p}={\bm \Gamma}(\lambda) {\bf p}.
\end{eqnarray}
Here the transition matrix is given by, 
\begin{equation}
{\bm \Gamma}(\lambda)
=
\left( 
\begin{array}{ccc}
-\Gamma_{ RL}-\Gamma_{ DL} & \Gamma_{ LR} e^{-i \lambda} & \Gamma_{ LD} \\
\Gamma_{ RL} e^{ i \lambda} &  -\Gamma_{ LR} -\Gamma_{DR} & \Gamma_{ RD} \\
\Gamma_{ DL} & \Gamma_{ DR} & -\Gamma_{LD}-\Gamma_{RD}
\end{array}
\right). 
\end{equation}
Figure~\ref{Fig:mc1} indicates six transitions with $\Gamma_{ij}$ between three charge states.  
Following the recipe of the full-counting statistics (FCS) of Bagrets and Nazarov \cite{Bagrets}, 
we introduced the counting field $\lambda$, which keeps track of the electrons transfered through the tunnel barrier between the two quantum dots (Fig.~\ref{Fig:mc1}). 
Then the probability distribution of the charge transfered through this barrier during the time $\tau$ is given by the Fourier transform
\begin{equation}
P_\tau(q)=\int_{-\pi}^\pi \frac{d\lambda}{2\pi}\, 
e^{-i\lambda q}{\cal Z}_\tau(\lambda),
\end{equation} 
where 
\begin{equation} 
{\cal Z}_\tau(\lambda)=p_L(\tau)+p_R(\tau)+p_D(\tau) ,
\end{equation} 
is the characteristic function. 
In the long time limit $\tau\gg I/e$, the characteristic function takes the form ${\cal Z}(\lambda)\sim e^{\tau{\cal F}(\lambda)}$,
where ${\cal F}(\lambda)$ is the eigenvalue of the matrix ${\bm \Gamma}(\lambda)$ with the largest real part.
The function ${\cal F}(\lambda)$ has to be found from the characteristic equation
\begin{eqnarray}
0&=&\det[ {\bf \Gamma}(\lambda) - {\cal F} \, {\bf I}]
 = {\cal F}^3 + K {\cal F}^2 + K' {\cal F} 
\nonumber\\ &&
+\,\Gamma_{ DR}\Gamma_{ RL}\Gamma_{ LD}({\rm e}^{i\lambda} - 1)
+\Gamma_{ DL} \Gamma_{ LR} \Gamma_{ RD}({\rm e}^{-i\lambda} - 1), 
\label{eigen}
\end{eqnarray}
where $K$ and $K'$ are parameters independent of the counting field,
\begin{eqnarray}
K = \sum_{i \neq j} \Gamma_{ij}, 
\;\;\;\;
K'= 
\sum_{i \neq j} 
\sum_{k} 
\Gamma_{i k} \Gamma_{k j}
+
\sum_{i \neq j} 
\sum_{k} 
\Gamma_{k i} \Gamma_{k j}/2 \, .
\end{eqnarray}
Without solving this equation we observe that ${\cal F}(\lambda)$
and hence ${\cal Z}(\lambda)$ in the long time limit satisfy the identity
\begin{eqnarray}
{\cal Z}(\lambda)={\cal Z}\left(-\lambda+i\frac{eV_S}{ k_{\rm B}T^*}\right),
\label{identity}
\end{eqnarray}
where the effective temperature $T^*$ is 
\begin{equation}
T^*=\frac{eV_S}{k_{\rm B} \ln w}, 
\label{T*}
\end{equation}
\begin{equation}
w
=
\frac
{\Gamma_{ DR}\Gamma_{ RL}\Gamma_{ LD}}
{\Gamma_{ DL} \Gamma_{ LR} \Gamma_{ RD}} \, . 
\label{w}
\end{equation}
Performing the inverse Fourier transformation of Eq. (\ref{identity}), we arrive
at the relation (\ref{pft}) for the distribution $P_\tau(q)$ with $T$ being replaced by $T^*$.

One can demonstrate \cite{our_paper} that the effective
temperature (\ref{T*}) is equal to the base temperature, 
$T^*=T$, when the tunneling rates $\Gamma_{ij}$ satisfy the detailed balance relation 
\begin{eqnarray}
\frac{\Gamma_{ij}}{\Gamma_{ji}} 
= 
\exp \left( \frac{\Delta_j-\Delta_i}{ k_{\rm B}T} \right), 
\end{eqnarray}
where $\Delta_j$ are the electrochemical potentials of the charge states of the DQD. 
In real experiments the QPC is biased and generates a non-equilibrium shot noise. 
Under these conditions the detailed balance is violated and,  as we will show later in Sec.~\ref{sec:be}, we have $T^*>T$. 
Although the canonical form of the FT (\ref{pft}) is violated in this case, the more general form (\ref{DS}) still holds. 
The reason is the underlying simplicity of the considered system, 
in which a forward transfer of a single electron occurs only through the following cycle of transitions 
$|D\rangle\to |L\rangle \to |R\rangle \to |D\rangle$ (see Fig.~\ref{Fig:mc1}). 
It enables us to introduce the macroscopic affinity $\ln w$ uniquely~\cite{Lebowitz,Andrieux,Andrieux2,Schnakenberg}. 
For a general system with more cycles in its state transition diagram, 
the FT holds only when the affinities for all cycles associated with current flow into/out of a particular lead coincide~\cite{Andrieux2,Schnakenberg}. 
Later in Sec.~\ref{sec:finite}, we will show that a detector with finite bandwidth leads to a violation of this condition.

\section{Effects of backaction and environments}
\label{sec:be}

Mesoscopic electron transport suffers from  environmental effects. 
In GaAs nanostructures, acoustic phonons strongly couple to electrons via the deformation potential and the piezoelectric coupling~\cite{Gasser}. 
For a QPC measurement, one cannot avoid the Coulomb interaction between the dots and QPC leads, which is marked by $C_{L}$ and $C_{R}$ in Fig.~\ref{Figsetups} a~\cite{Gustavsson2,Aguado,Hashisaka}. 
This interaction is unwanted and causes a measurement backaction, 
since the nonequilibrium QPC shot noise leads to QD level fluctuations. 
In addition, the external circuit acts as the electromagnetic environment, 
which further affects both of the DQD and QPC. 

Such environmental effects can be accounted for by the so-called $P(E)$-theory~\cite{Ingold} and more systematically by using the real-time diagrammatic technique~\cite{Koenig}. 
The phonon and electromagnetic environments, as well as 
the nonequilibrium QPC current noise generate QD level fluctuations, 
$\delta V_L$ and  $\delta V_R$. 
Then the tunnel rates connecting the three charge states are modified as, 
\begin{eqnarray}
\Gamma_{\! LR}
\! &=& \!
\pi \, |T_C|^2 \, P_C(E_L-E_R)
\, ,
\\
\Gamma_{\! DL} 
\! &=& \!
\Gamma_{\! R} 
\int \!\! d \omega 
f(E_D-E_L-V_{\rm S}/2-\omega) \, P_R(\omega) \, ,
\\
\Gamma_{\! RD}
\! &=& \!
\Gamma_{\! L} 
\int \!\! d \omega 
f(\omega-E_R+V_{\rm S}/2+E_D) \, P_L(\omega) \, .
\end{eqnarray}
($\Gamma_{RL}$, 
$\Gamma_{LD}$ and 
$\Gamma_{RD}$ are given in a similar manner). 
Here $T_C$ is a tunnel matrix element describing the central barrier. 
The total energies of the charge states $E_j$ include the electrostatic energy. 
The tunnel rates $\Gamma_{\! DL}$ and $\Gamma_{\! RD}$ are also
affected by the thermal broadening of the reservoir levels through the Fermi distribution, 
$f(\omega) \!=\! 1/(1 \!+\! {\rm e}^{\omega/k_{\rm B} T})$. 

The Fourier transform of the correlation function, 
$P_j(\omega)
\!=\! 
\int \! dt \, {\rm e}^{i \omega t} P_j(t)/(2 \pi)$ 
induces additional broadening. 
It is determined by the fluctuating dot potentials, 
$\varphi_{L/R}(t) \!=\! \int^t \! d t' \delta V_{L/R}(t')$ 
and 
$\varphi_C \!=\! \varphi_{R} \!-\! \varphi_{L}$, 
as 
\begin{equation}
P_j
\!=\!
\left \langle \!
{\rm e}^{ i \hat{\varphi}_j(t)}
{\rm e}^{-i \hat{\varphi}_j(0)} 
\! \right \rangle
\! = \!
\exp \! \left[ \! 
\int \!\! d \omega
\frac{
S_{\delta V j} (\omega) 
( e^{-i \omega t} \!-\! 1)
}{\omega^2} 
\right],
\label{ppc}
\end{equation}
where the correlation function for the level fluctuations 
is determined by the properties of the environment.

For  acoustic phonons the correlation function takes the following form 
\begin{eqnarray}
S^{\rm ph}_{\delta V j} (\omega)
\!=\!
\frac{A^{\rm ph}_j(\omega)}
{1 \!-\! {\rm e}^{-\omega/k_{\rm B} T}}
\label{eqn:sph}
\end{eqnarray}
with super-ohmic, 
$A^{\rm ph}_C \! \propto \! \omega^3$, 
or ohmic, 
$A^{\rm ph}_{L/R} \! \propto \! \omega$, 
phonon spectral functions. 
Though phonons cause some additional broadening, for the experiment considered with low measurement current,
$\langle I_{\rm QPC} \rangle \! \approx \! 12 {\rm nA}$,  
the heating effect is negligible~\cite{Gasser}. 
As long as the lattice and electronic systems are isothermal 
the detailed balance holds, and as a consequence the FT, Eq.~(\ref{pft}), is satisfied. 

The situation changes when the environment itself is out of equilibrium. 
Such an environment is generated by the QPC current fluctuations $S_I^{\rm QPC}$, 
which give rise to the voltage fluctuation spectrum 
\begin{eqnarray}
S^{\rm QPC}_{\delta V j} (\omega) 
=
\kappa_j |Z_t(\omega)|^2 S_I^{\rm QPC}(\omega) .
\label{eqn:sqpc}
\end{eqnarray}
($\kappa_{L/R} \!=\! 1$, $\kappa_{C} \!=\! 4$). 
The non-symmetrized current noise of the QPC is given by
\begin{eqnarray}
S_I^{\rm QPC}
\! = \!
\frac{2}{R_{\rm K}}
\left[
\sum_\pm
\frac{
{\cal T}_{\rm QPC}(1 \!-\! {\cal T}_{\rm QPC})
(\omega \pm V_{\rm QPC})
}{
1-{\rm e}^{-(\omega \pm V_{\rm QPC})/k_{\rm B} T}
}
+
\frac{2 \, {{\cal T}_{\rm QPC}}^2 \, \omega}{1-{\rm e}^{-\omega/k_{\rm B} T}}
\right]
.
\end{eqnarray}
The impedance $Z_t(\omega) \!=\! 1/(i \omega \, \bar{C} \!+\! 1/\bar{R})$ 
characterizes the capacitive coupling between the QPC and the 
dot-level fluctuations $\delta V_r$. 
Here
$\bar{R}$
is written with the QPC resistance $R_{\rm QPC}$ as 
$\bar{R}
\!=\! 
R_{\rm QPC}/[1 \!+\! \bar{C} (C_L^{-1} \!+\! C_R^{-1})]$. 
The capacitance is 
$
\bar{C}
\!=\!
(3 C_0 \!+\! C_G)/2
$,
where the capacitances $C_0$ and $C_G$ characterizes the coupling between dots, leads and gate electrodes (Fig.~\ref{Figsetups} a). 
For the experiment~\cite{Fujisawa}, the QPC transparency for each spin is estimated as 
${\cal T}_{\rm QPC}
\!=\!
R_{\rm K}/(2 R_{\rm QPC})
\! \approx \!
0.19$
and 
$V_{\rm QPC} \!=\! 0.8 {\rm mV}$. 
Equation~(\ref{eqn:sqpc}) is reduced to the equilibrium form Eq.~(\ref{eqn:sph}),
when $V_{\rm QPC}=0$ and the detailed balance is satisfied. 
However, for $V_{\rm QPC} \neq 0$, the detailed balance and thus the FT, Eq.~(\ref{pft}), are violated. 
Note that the violation is not contradict to the FT for the total DQD and QPC system, Eq.~(\ref{P2}).

Generally the electromagnetic environment suppresses the phase correlations. 
In the long-time limit, it decays exponentially 
$P_j(t) \! \approx \! \exp(-\Gamma^{\rm QPC}_j t/2)$. 
For realistic parameters, 
$\bar{C} \!=\! 5 {\rm fF}$ and $\bar{C}(C_L^{-1} \!+\! C_R^{-1}) \!=\! 0.02$, 
the decay rate is rather big, 
$
\Gamma^{\rm QPC}_C/2
\!=\! 
\pi |Z_t(0)|^2 S_I(0) 
\! \approx \! 
50 \mu{\rm eV}
$. 
Even for the ideal case, i.e. there is no capacitive coupling between the QPC and the DQD, $C_{L/R} \!=\! 0$, the correlation function $P_j$ decays exponentially. 
It is because of an {\it intrinsic} backaction often discussed in the context of the weak measurement~\cite{Averin}. 
For the ideal case, the decay rate is, 
\begin{equation}
\frac{\Gamma^{\rm QPC}_C}{2}
\approx
- \frac{V_{\rm PC} \, t}{\pi}
\ln \!
\left(
\sqrt{{\cal T}_{| L \rangle}}
\sqrt{{\cal T}_{| R \rangle}}
 \!+\! 
\sqrt{1 \!-\! {\cal T}_{| L \rangle}}
\sqrt{1 \!-\! {\cal T}_{| R \rangle}}
\right),
\end{equation}
in the limit of 
$t \! \to \! \infty$ 
for 
$T \! \ll \! V_{\rm QPC}$. 
Here ${\cal T}_{| s \rangle}$ means the transmission probability through the QPC when the DQD is in the state $|s \rangle$. 
However, for the experiment~\cite{Fujisawa}, $I_{| L/R \rangle} \! \sim \! 12 \pm 0.1 {\rm nA}$, 
and thus we estimate $\Gamma^{\rm QPC}_C/2 \! \sim \! 2.1 {\rm nV}$, which is negligible.

\section{Effect of finite detector bandwidth}
\label{sec:finite}

\begin{figure}[th]
\begin{center}
\includegraphics[width=.5 \columnwidth]{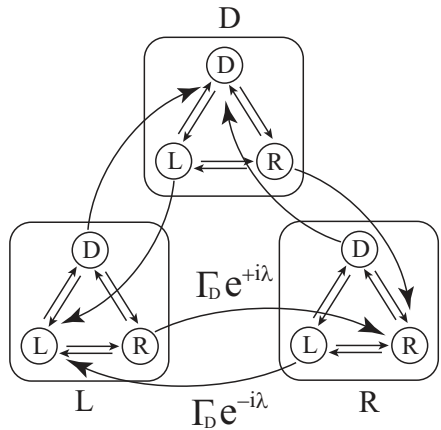}
\caption{
State transition diagram for the double dot and the QPC detector with finite bandwidth. 
Circles represent the double-dot states and arrows show the directions of the transitions. 
The additional states representing the detector states (squares) are introduced. 
Within each square, transitions between all the dot states are possible. 
Between the 3 detector states, one-way transition, the relaxation from the `false' detector state to the `true' detector state, occurs. 
} 
\label{Fig:mc2}
\end{center}
\end{figure}

In this section we discuss the effect of the finite bandwidth of the measurement device on the FT. 
If the detector bandwidth is finite, the QPC current does not follow the switching between the DQD charge states immediately. 
Naaman and Aumentado\cite{naaman} proposed to describe such a system by doubling the number of states. 
In our system the DQD switches between three states 
$|D\rangle$, $|L\rangle$, $|R\rangle$, 
and the QPC current takes three values corresponding to those states 
$|D \rangle_{\rm QPC}$, $|L \rangle_{\rm QPC}$, $|R \rangle_{\rm QPC}$. 
Then, we should describe the system by 9 states 
$|r \rangle |r' \rangle_{\rm QPC}$ 
($r,r'=L,R,D$), 
and we have to consider a vector of 9 occupation probabilities 
\begin{equation} 
{\bf p}^T=(p_{LL},p_{RL},p_{DL},p_{LR},p_{RR},p_{DR},p_{LD},p_{RD},p_{DD}),
\end{equation}
where the first index refers to the state of the DQD and the second one
to the value of the QPC current. 
As shown in Fig. \ref{Fig:mc2}, the detector will always change to the state corresponding to the dot-state. 
We model the fact that the detector needs a finite time for this switching, by introducing the detector rate $\Gamma_D$. 
For an ideal detector, $\Gamma_D\rightarrow \infty$,  only the states are $p_{LL}$, $p_{RR}$ and $p_{DD}$ occur. 

Because of the one-way transitions between detector states
the considered system is outside the class discussed in Ref.~\cite{Andrieux2}. 
In order to describe the experiment and the consequences for the FT we calculate the cumulant generating function. 
We introduce the master equation of the total system, 
\begin{equation}
\partial_t {\bf p}={\bm M}(\lambda) \, {\bf p}, 
\label{master_eq_D}
\end{equation}
where the transition matrix is a $9 \times 9$ matrix, 
\begin{equation}
{\bm M}(\lambda)=
\left(
\begin{array}{ccc}
{\bm \Gamma}(0)-{\bm \Gamma}_{\!\!D 2}-{\bm \Gamma}_{\!\!D 3} & {\bm \Gamma}_{\!\!D 1} e^{-i \lambda} & {\bm \Gamma}_{\!\!D 1} \\
{\bm \Gamma}_{\!\!D 2} e^{i \lambda} & {\bm \Gamma}(0)-{\bm \Gamma}_{\!\!D 1}-{\bm \Gamma}_{\!\!D 3} & {\bm \Gamma}_{\!\!D 2} \\
{\bm \Gamma}_{\!\!D 3} & {\bm \Gamma}_{\!\!D 3} & {\bm \Gamma}(0)-{\bm \Gamma}_{\!\!D 1}-{\bm \Gamma}_{\!\!D 2} \\
\end{array}
\right), 
\end{equation}
with sub-matrices  given by
\begin{equation}
{\bm \Gamma}_{\!\!D i}={\Gamma}_{\! D} 
\left(
\begin{array}{ccc}
\delta_{i1} & 0 & 0 \\
0 & \delta_{i2} & 0 \\
0 & 0 & \delta_{i3} 
\end{array}
\right) \, .
\end{equation}
Note that the counting field is associated with the rate $\Gamma_D$. I.e., in the present model the 
switching of the QPC current are counted and not the transitions in the DQD system. 
Thus the model provides information about the experimentally accessible statistics of the detector 
rather than that of the DQD, which is not directly measurable. 
The model outlined above fits the measured higher cumulants for the single-dot case quite accurately~\cite{flindt,Gustavsson2}.

The FCS cumulant generating function 
$\cal F$ 
is obtained by solving the the characteristic equation for the eigenvalue
\begin{eqnarray}
0 &=& \det|{\bm M}(\lambda)-{\cal F} \, {\bf I}|
\nonumber\\ 
&=&
\det|{\bm M}(0)-{\cal F} \, {\bf I}|
+
\Gamma_+({\cal F})\left(e^{i\lambda}-1\right)
+
\Gamma_-({\cal F})\left(e^{-i\lambda}-1\right),
\end{eqnarray}
where $\Gamma_\pm$ are factorized as, 
\begin{equation}
\Gamma_+({\cal F}) 
=
\Gamma_D^6
(x \, \Gamma_{DL}+\Gamma_{DL}^*)
(x \, \Gamma_{LR}+\Gamma_{LR}^*)
(x \, \Gamma_{RD}+\Gamma_{RD}^*), 
\end{equation}
\begin{equation}
\Gamma_-({\cal F}) 
=
\Gamma_D^6
(x \, \Gamma_{LD}+\Gamma_{LD}^*)
(x \, \Gamma_{RL}+\Gamma_{RL}^*)
(x \, \Gamma_{DR}+\Gamma_{DR}^*). 
\end{equation}
We introduced $x={\cal F}/\Gamma_D$ and the corrected tunnel rates
\begin{eqnarray}
\Gamma_{ij}^*
=
\Gamma_{ij} \, 
\left(
1+\frac{\Gamma_{ik}+\Gamma_{jk}}{\Gamma_D}
\right) 
+ 
\frac{\Gamma_{ik} \Gamma_{kj}}{\Gamma_D}, 
\;\;
(k \neq i,j). 
\label{eq:Gijcorrected}
\end{eqnarray}

In order to check whether the FT is satisfied, we consider the following ratio,
generalizing $w$ of the ideal case Eq.~(\ref{w}), 
\begin{eqnarray}
w^*=\frac{\Gamma_+({\cal F})}{\Gamma_-({\cal F})}.
\label{R0}
\end{eqnarray}
The FT has to be exact if this ratio does not depend on ${\cal F}$. 
It is obvious that generally this is not the case and therefore the FT is violated. 
However, in two limits, for a fast detector 
$\Gamma_D \gg \Gamma_{ij}$ 
and for a slow detector 
$\Gamma_D \ll \Gamma_{ij}$, 
we are able to show analytically that the FT holds. 
In the former case, $\Gamma_D\gg\Gamma_{ij}$, 
we expand $w^*$ in powers of $1/\Gamma_D$. 
Since ${\cal F}_D \sim \Gamma_{ij}$, we arrive at the following result
\begin{equation}
w^* = 
w
+
\frac{1-w}{\Gamma_D}
\left(  \frac{\Gamma_{LD}\Gamma_{DR}}{\Gamma_{LR}} 
+ \frac{\Gamma_{RL}\Gamma_{LD}}{\Gamma_{RD}}
+ \frac{\Gamma_{DR}\Gamma_{RL}}{\Gamma_{DL}}\right)
+{\cal O}\left(\frac{\Gamma_{ij}^2}{\Gamma_D^2}\right). 
\label{R1}
\end{equation}
\begin{figure}[t]
\begin{center}
\includegraphics[width= 8 cm]{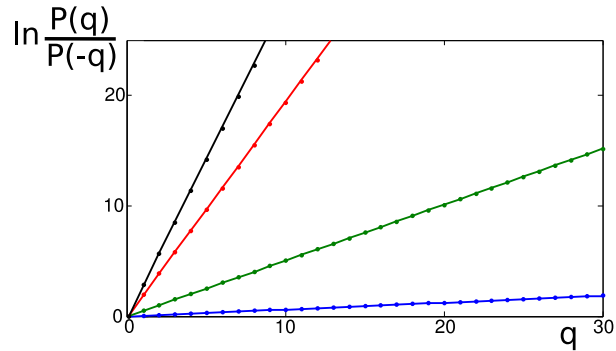}
\caption{The ratio of the probability of transfered charge as a function of the number 
of charges in the long-time limit. The solid lines are the results given by the analytical expression for the effective temperature [see Eq.~(\ref{T*1})], the dots are from fully numerical calculations. 
The different colors correspond to different rates for the detector bandwidth: 
(black) $\Gamma_D=100\,{\rm kHz}$, (red) $\Gamma_D=10\,{\rm kHz}$, (green) $\Gamma_D=1\,{\rm kHz}$, (blue) $\Gamma_D=0.1\,{\rm kHz}$. 
The rates for the transitions on the dot are the same as used in Ref. \cite{our_paper} and are of the order of $~1{\rm kHz}$:
$\Gamma_{DR} \!=\! 4 {\rm kHz}$, 
$\Gamma_{RD} \!=\! 0.3 {\rm kHz}$, 
$\Gamma_{DL} \!=\! 1 {\rm kHz}$, 
$\Gamma_{LD} \!=\! 1.5 {\rm kHz}$, 
$\Gamma_{LR} \!=\! 1.7 {\rm kHz}$, 
and 
$\Gamma_{RL} \!=\! 1.8 {\rm kHz}$. 
}
\label{Fig:ET}
\end{center}
\end{figure}
We observe that in the lowest and next to  lowest orders in the parameter $\Gamma_{ij}/\Gamma_D$ the ratio $w^*$ does not depend on ${\cal F}$ and thus the FT holds. 
Comparing Eqs. (\ref{eq:Gijcorrected}) and (\ref{R1}) we note that within the accuracy of our approximation, i.e. up to the terms  $\sim 1/\Gamma_D$, the ratio $w^*$ may be written in the same form as for an ideal detector but with modified tunnel rates, 
\begin{eqnarray}
w^*
\approx 
\frac{\Gamma_+(0)}{\Gamma_-(0)}
=
\frac{\Gamma_{DR}^*\Gamma_{RL}^*\Gamma_{LD}^*}{\Gamma_{LR}^*\Gamma_{RD}^*\Gamma_{DL}^*}, 
\label{R}
\end{eqnarray}
and  effective temperature 
\begin{eqnarray}
T^*=\frac{eV_S}{k_B \ln w^*}.
\label{T*1}
\end{eqnarray}

In the opposite limit of a slow detector,
$\Gamma_D\ll\Gamma_{ij}$, 
one can show that
${\cal F} \sim \Gamma_D \ll \Gamma_{ij}$
 and therefore one can put ${\cal F}=0$ in Eq. (\ref{R0}). 
Surprisingly, this means that  Eqs. (\ref{R}-\ref{T*1})  also become valid in the opposite limit, that of a very slow detector. 
Numerical calculations of ${\cal F}$ confirm these results, but as expected ${\cal F}$ does not have the right symmetries for intermediate values of $\Gamma_D$. 

In Fig.~\ref{Fig:ET} we compare our result for the effective temperature with numerical results. 
Varying the detector bandwidth over four orders of magnitude we see that our expression for the effective temperature Eq.~(\ref{T*1}) provides good fits for rather wide range of parameters of $q$ and $\Gamma_D$. 
This means that although formally the detector model introduced by Naaman and Aumentado violates the FT, practically, the finite bandwidth effect can be accounted for simply by the effective temperature. 
Thus Eqs. (\ref{R}-\ref{T*1}) should describe the experiment  reasonably well, regardless of the value of $\Gamma_D$.

To conclude, in this section we have demonstrated that a finite detector bandwidth
in general distorts the measured statistics of the charge transfer and leads
to the formal violation of FT. However the effect of the 
 finite bandwidth can be accounted for by using the expression~(\ref{T*1}) for the 
 effective temperature.

\section{Tunneling rates in the experiment}
\label{sec:wtd}

Let us now discuss how to determine the values of $\Gamma_{ij}$ and $\Gamma_D$ from  experimental data. The standard way of doing this is to generate  dwell time histograms for every current state. 
Let us consider the QPC current state corresponding to the DQD state $|L\rangle$.
One should count how many times during sufficiently long observation time the QPC current has switched from the state $| L \rangle$ to either state $|R\rangle$ or $|D\rangle$ within the time interval from $\tau$ and $\tau+\Delta\tau$, where $\Delta\tau$ is  sufficiently short. Denoting the corresponding numbers $\Delta N_{L\to R}$ and $\Delta N_{L\to D}$, one gets the histograms plotting the values $\Delta N_{L\to R,\;L\to R}(\tau)$ as a function of time $\tau$.  At sufficiently long time $\tau$ the numbers $\Delta N_{L\to R,\;L\to R}(\tau)$ decay in time exponentially,
\begin{eqnarray}
\Delta N_{L\to R}(\tau) = K_{RL}e^{-\Gamma_{L}^*\tau},\;\;
\Delta N_{L\to D}(\tau) = K_{DL}e^{-\Gamma_{L}^*\tau}.
\label{histogram} 
\end{eqnarray}
The parameters $\Gamma_L^*$, $K_{RL}$ and $K_{DL}$ are extracted
by fitting the histograms. 
 
In the case of a fast detector, $\Gamma_D\gg\Gamma_{ij}$ one can easily
express the DQD tunneling rates $\Gamma_{RL}$ and $\Gamma_{DL}$ in terms
of the parameters $\Gamma_L^*$, $K_{RL}$, $K_{DL}$. Indeed,
the theory in this case predicts $\Gamma_L^*=\Gamma_{RL}+\Gamma_{DL}$ and $K_{RL}/K_{DL}=\Gamma_{RL}/\Gamma_{DL}$
and, therefore
\begin{eqnarray}
\Gamma_{RL}=\frac{K_{RL}}{K_{RL}+K_{DL}}\Gamma_L^*,\;\;
\Gamma_{DL}=\frac{K_{DL}}{K_{RL}+K_{DL}}\Gamma_L^*.
\end{eqnarray}
The remaining four rates are determined analogously.

In the following, we will present our analysis in detail. 
Suppose at time $\tau=0$ the QPC current has switched to the state $L$.
Since the state of the DQD remains unknown, the total occupation probability
of such  a state is given by the sum $p_L=p_{LL}+p_{RL}+p_{DL}$.
To find out how this probability decays in time we have to solve the equation for the probabilities 
${\bf p}^T=(p_{LL},p_{RL},p_{DL})$, 
\begin{equation}
\partial_t {\bf p}
=
({\bf \Gamma}(0)-{\bf \Gamma}_{\!\! D2}-{\bf \Gamma}_{\!\! D3})
\, {\bf p} \, . 
\end{equation}
This equation is a sub-block of a more general equation (\ref{master_eq_D})
where only the outgoing processes from the QPC state $L$ are kept.
For a large number of events the histograms should converge
to the following expressions
\begin{eqnarray}
\Delta N_{L\to R}(\tau) = N\Gamma_D p_{RL} ,\;\;
\Delta N_{L\to D}(\tau) = N\Gamma_D p_{DL},
\end{eqnarray} 
where $N$ is the normalization factor.
Solving the differential equation and considering the long-time limit
we arrive at Eqs. (\ref{histogram}) with
\begin{eqnarray}
\Gamma_L^* = \Gamma_{RL}+\Gamma_{DL} 
-\frac{\Gamma_{RL}\Gamma_{LR}+\Gamma_{DL}\Gamma_{LD}}{\Gamma_D} +{\cal O}\left(\frac{1}{\Gamma_D^2}\right)
\end{eqnarray}
and
\begin{eqnarray}
\frac{K_{RL}}{K_{DL}}
=
\frac{\Gamma_{RL}^*}{\Gamma_{DL}^*}+{\cal O}\left(\frac{1}{\Gamma_D^2}\right).
\end{eqnarray}
Thus the ratio of the prefactor is sufficient to estimate the effective temperature (\ref{T*1}).

One can also obtain the full time-dependence of $\Delta N_{L\to R}(\tau)$ and $\Delta N_{L\to D}(\tau)$.
To this end we put $p_{LL}(0)=1,p_{RL}(0)=p_{DL}(0)=0$ and in the limit $\Gamma_D\gg\Gamma_{ij}$
obtain the  result
\begin{eqnarray}
\Delta N_{L\to R}=
\frac{N\Gamma_D[\Gamma_{RL}(\Gamma_D+\Gamma_{LD}+\Gamma_{RD}-\Gamma_L^*)
+\Gamma_{RD}\Gamma_{DL}]\big[e^{-\Gamma_L^*\tau}-e^{-\Gamma_D\tau}\big]}
{(\Gamma_D+\Gamma_{LR}+\Gamma_{D2}-\Gamma_L^*)(\Gamma_D+\Gamma_{LD}+\Gamma_{RD}-\Gamma_L^*)-\Gamma_{DR}\Gamma_{RD}},
\nonumber\\
\Delta N_{L\to D}=
\frac{N\Gamma_D[\Gamma_{DL}(\Gamma_D+\Gamma_{LR}+\Gamma_{DR}-\Gamma_L^*)
+\Gamma_{DR}\Gamma_{RL}]\big[e^{-\Gamma_L^*\tau}-e^{-\Gamma_D\tau}\big]}
{(\Gamma_D+\Gamma_{LR}+\Gamma_{D2}-\Gamma_L^*)(\Gamma_D+\Gamma_{LD}+\Gamma_{RD}-\Gamma_L^*)-\Gamma_{DR}\Gamma_{RD}}.
\nonumber
\end{eqnarray} 
With the aid of these expressions one can extract $\Gamma_D$ from the experimental data.

\section{Summary}
\label{sec:sum}

We have discussed the possibility of experimental verification of the
FT (\ref{pft}) in single-electron counting experiments. Although the experiments
 allow  testing the FT, the properties of the
detector turn out to be very important. 
We have shown that in a generic system
of a double quantum dot the FT (\ref{pft}) is robust against the backaction
of a QPC detector in the sense that the effect of the latter can be absorbed in an effective temperature (\ref{T*}).
We also investigated the influence of the finite bandwidth of the detector. 
We found that finite bandwidth results in the distortion of the signal properties and formally cannot
be reduced to the renormalization of the effective temperature in Eq. (\ref{pft}).
However, in practice these deviations turn out to be small, and Eq. (\ref{pft})
with modified temperature (\ref{T*1}) should describe the data quite well,
even if they are obtained with a slow detector. 
We hope the experimental test of the FT in the single-electron transport~\cite{our_paper}, as well as that in the Aharonov-Bohm ring~\cite{Nakamura}, would stimulate the development in the nonequilibrium statistical physics and the mesoscopic quantum physics.

\section{Acknowledgement}
We thank M.~Hettler, K.~Kobayashi, and K.~Saito for valuable discussions.
This work has been supported by Strategic International Cooperative Program 
of the Japan Science and Technology Agency (JST) and 
by the German Science Foundation (DFG).

\end{document}